\documentclass[twocolumn,prl,showpacs,preprintnumbers,amsmath,amssymb]{revtex4}

\usepackage{graphicx}

\begin{document}

\title{$B=3$ Tetrahedrally Symmetric Solitons in the Chiral Quark Soliton Model}

\author{Nobuyuki Sawado, Noriko Shiiki}
\affiliation{Department of Physics, Tokyo University of Science, Noda, Chiba 278-8510, Japan}

\date{\today}

\begin{abstract}

 In this paper, $B=3$ soliton solutions with tetrahedral symmetry are obtained
numerically in the chiral quark soliton model using the rational map ansatz. 
The solution exhibits a triply degenerate bound spectrum of the quark orbits 
in the background of tetrahedrally symmetric pion field configuration. 
The corresponding baryon density is tetrahedral in shape. Our numerical 
technique is independent on the baryon number and its application to 
$B \geq 4$ is straightforward.

\end{abstract}


\maketitle


 The Chiral Quark Soliton Model (CQSM) was developed in 1980's as an effective 
theory of QCD interpolating between the Constituent Quark Model and Skyrme 
Model \cite{diakonov,cqsm}. In the large $N_{c}$ limit, these models are identical 
\cite{manohar}. 

The CQSM is derived from the instanton 
liquid model of the QCD vacuum and incorporates the non-perturbative 
feature of the low-energy QCD, spontaneous chiral symmetry breaking.  
The vacuum functional is defined by;

\begin{eqnarray}
	{\cal Z} = \int {\cal D}\pi{\cal D}\psi{\cal D}\psi^{\dagger}\exp \left[ 
	i \int d^{4}x \, \bar{\psi} \left(i\!\!\not\!\partial
	- mU^{\gamma_{5}}\right) \psi \right]	 \label{vacuum_functional}
\end{eqnarray} 
where the SU(2) matrix
\begin{eqnarray}
	U^{\gamma_{5}}= \frac{1+\gamma_{5}}{2} U + \frac{1-\gamma_{5}}{2} U^{\dagger} 
	\,\,\,{\rm with} \,\,\,\,
	U=\exp \left( i \vec{\tau} \!\cdot\! \vec{\pi}/f_{\pi} \right) \nonumber
\end{eqnarray}
describes chiral fields, $\psi$ is quark fields and $m$ is the dynamical 
quark mass. $f_{\pi} $ is the pion decay constant and experimentally 
$f_{\pi} \sim 93 {\rm MeV}$. 
Since our concern is the tree-level pions and one-loop quarks according 
to the Hartree mean field approach, the kinetic term of the pion fields which 
gives a contribution to higher loops can be neglected. 
Due to the interaction between the valence quarks and the Dirac sea, 
soliton solutions appear as bound states of quarks in the background of self-consistent 
mean chiral field. $N_{c}$ valence quarks fill the each bound state to form a baryon. 
The baryon number is thus identified with the number of bound states filled by 
the valence quarks \cite{kahana}. 

For $B=1$ and $2$, the spherically symmetric soliton \cite{meissner,reinhardt,wakamatsu} 
and the axially symmetric soliton \cite{sawado} were found respectively. Upon 
quantization, the intermediate states of nucleon and deuteron 
between the Constituent Quark Model and Skyrme Model were obtained. 

The vacuum functional in Eq.(\ref{vacuum_functional}) can be integrated 
over the quark fields to obtain the effective action 

\begin{eqnarray}
	S_{{\rm eff}}[U]&=&-iN_{c}{\rm lndet}\left(i
	\!\!\not\!\partial - mU^{\gamma_{5}}\right)\label{effective_action1}\\
	&=&-\frac{i}{2}N_{c}{\rm Spln}D^{\dagger}D
	\label{effective_action2}
\end{eqnarray}
where $D=i\!\!\not\!\partial - mU^{\gamma_{5}}$. 
This determinant is ultraviolet divergent and must be 
regularized. Using the proper-time regularization scheme, we can write 

\begin{widetext}
\begin{eqnarray}
	S^{{\rm reg}}_{{\rm eff}}[U]=\frac{i}{2}N_{c}
	\int^{\infty}_{1/\Lambda^2}\frac{d\tau}{\tau}{\rm Sp}\left(
	{\rm e}^{-D^{\dagger}D\tau}-{\rm e}^{-D_{0}^{\dagger}D_{0}\tau}\right) 
	=\frac{i}{2}N_{c}T\int^{\infty}_{-\infty}\frac{d\omega}{2\pi}
	\int^{\infty}_{1/\Lambda^2}\frac{d\tau}{\tau}{\rm Sp}\left[{\rm e}
	^{-\tau (H^2+\omega^2)}-{\rm e}^{-\tau (H_{0}^2+\omega^2)}\right] 
	\label{regularised_action}
\end{eqnarray}
\end{widetext}
where $T$ is the Euclidean time separation, $\Lambda$ is a cut-off 
parameter evaluated by the condition that the derivative expansion of 
Eq.(\ref{effective_action1}) reproduces the pion kinetic term with the 
correct coefficient ${\it i.e.}$
\begin{eqnarray}
	f_{\pi}^2=\frac{N_{c}m^2}{4\pi^2}\int^{\infty}_{1/\Lambda^2} 
	\frac{d\tau}{\tau}{\rm e}^{-\tau m^2}
	\,\, , \label{cutoff_parameter}
\end{eqnarray}
and $H$ is the Dirac one-quark Hamiltonian defined by
\begin{eqnarray}
	H=\frac{\vec{\alpha}\!\cdot\!\vec{\nabla}}{\it{i}}+\beta m 
	U^{\gamma_{5}}\,\,. \label{}
\end{eqnarray}
$D_{0}\equiv D(U=1)$ and $H_{0}\equiv H(U=1)$ correspond to 
the vacuum sectors. 

At $T \rightarrow \infty$, we have ${\rm e}^{iS_{{\rm eff}}}
\sim  {\rm e}^{-iE_{sea}T}$. Integrating over $\omega$ 
in (\ref{regularised_action}) and constructing a complete set of 
eigenstates of $H$ with 
\begin{eqnarray}
	H|\nu\rangle = \epsilon_{\nu}|\nu\rangle\,\, ,\,\,\,
	H_{0}|\nu\rangle^{(0)} = \epsilon^{(0)}_{\nu}|\nu\rangle^{(0)}
	\,\, , \label{eigen_equation}
\end{eqnarray}
one obtains the sea quark energy \cite{meissner2}
\begin{widetext}
\begin{eqnarray}
	E_{sea}[U]=\frac{1}{4\sqrt{\pi}}N_{c}\int^{\infty}_{1/\Lambda^2}
	\frac{d\tau}{\tau^{3/2}}\left(\sum_{\nu}{\rm e}^{-\tau\epsilon_{\nu}^2}
	-\sum_{\nu}{\rm e}^{-\tau\epsilon^{(0)2}_{\nu}}\right)\,\, .
	\label{energy_sea} \label{enegy_sea}
\end{eqnarray}
In the Hartree picture, the baryon states are the quarks occupying all 
negative Dirac sea and valence levels. Hence, if we define the total soliton energy 
$E_{total}$, the valence quark energy should be added;
 \begin{eqnarray}
	E_{total}[U]=N_{c}\sum_{i}E_{val}^{(i)}[U]+E_{sea}[U]\,\,. 
	\label{energy_total}
\end{eqnarray}
where $E_{val}^{i}$ is the valence quark contribution to the $i$ th baryon.

The baryon density $\langle b_{0}\rangle$ for the baryon number $B$ soliton is defined by 
the zeroth component of the baryon current \cite{reinhardt}; 
\begin{eqnarray}
	\langle b_{0}\rangle &=& \frac{1}{N_{c} B}\langle\bar{\psi}\gamma_{0}
	\psi\rangle  
	= \frac{1}{N_{c} B}\left[\sum_{\nu}\bigl(n_\nu\theta(\epsilon_{\nu}) +{\rm sign}(\epsilon_{\nu})
	{\cal N} (\epsilon_{\nu})\bigr)\langle \nu|\vec{r}\rangle\langle \vec{r}|\nu\rangle 
	-\sum_{\nu}{\rm sign}(\epsilon_{\nu}^{(0)}){\cal N}(\epsilon_{\nu}^{(0)})	
	\langle \nu|\vec{r}\rangle^{(0)} \langle\vec{r}|\nu\rangle^{(0)}
	\right] 	 \label{baryon_density}
\end{eqnarray}
\end{widetext}
where
\begin{eqnarray*}
{\cal N}(\epsilon_{\nu})=-\frac{1}{\sqrt{4\pi}}\Gamma\left(\frac{1}{2},
\left(\frac{\epsilon_\nu}{\Lambda}\right)^2\right)
\end{eqnarray*}
and $n_\nu$ is the valence quark occupation number.  

\begin{figure*}
\includegraphics[height=9cm,width=14cm]{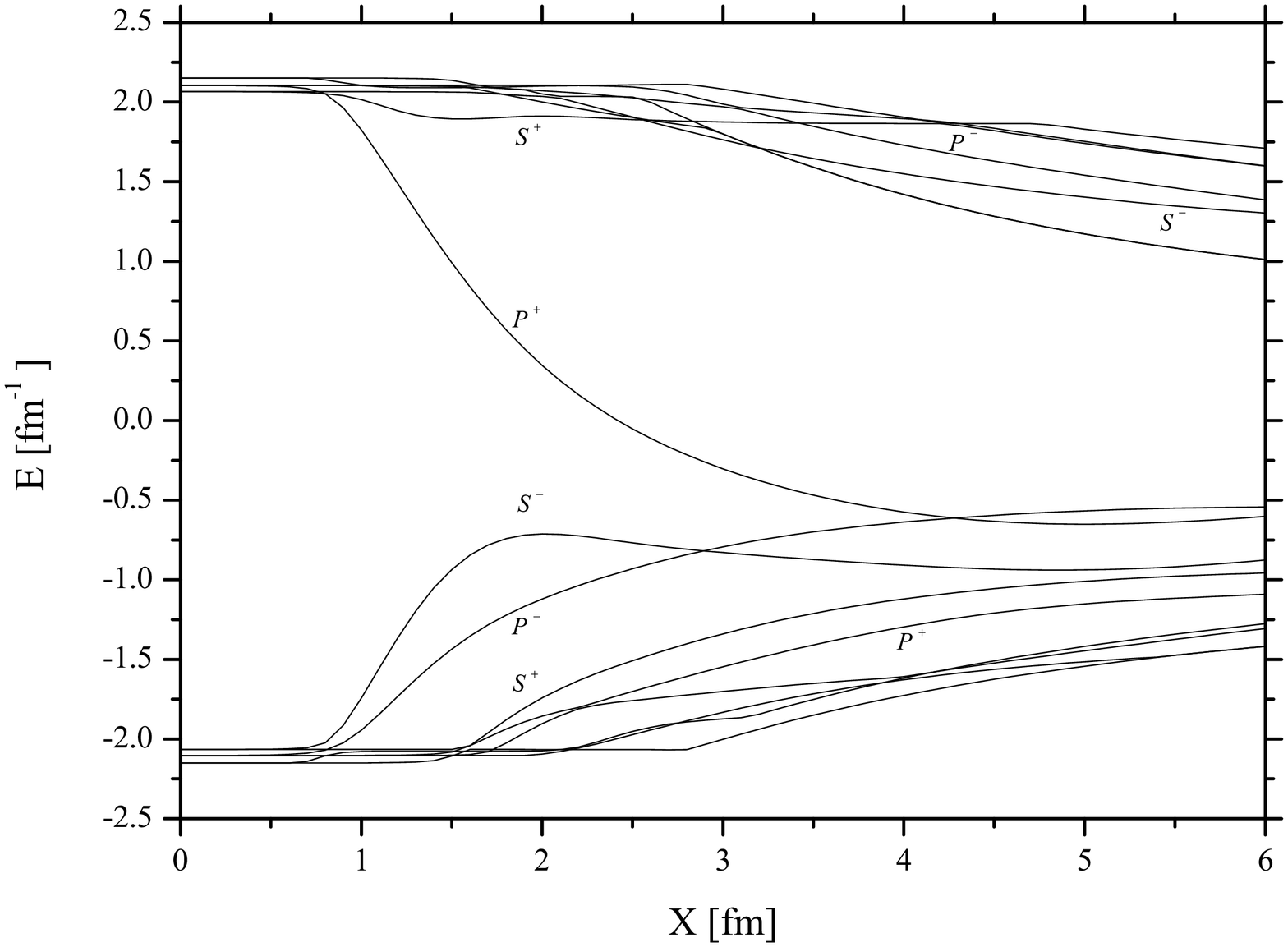}
\caption{\label{fig:spectrum} Spectrum of the quark orbits are illustrated as a function of the 
``soliton size'' $X$. Trial profile function is defined by 
$f(r)=-\pi+\pi r/ X$ for $X \le r$ and $f(r)=0$ for $X > r$.
The state coupled with $(K,M)=(0,0)$ is denoted as $S$ and $(K,M)=(1,M)$ as $P_M$ 
($M=\pm1,0$) with parity $\pm$. The energy of the three $P$ states are all degenerate. }
\end{figure*}

For $B=3$, it is expected the solution to have a tetrahedral symmetry from the 
study of the Skyrme model \cite{braaten}. Therefore, we shall impose the same 
symmetry on the chiral fields using the rational map ansatz. 
According to the ansatz, the chiral field can be expressed as \cite{manton} 

\begin{eqnarray}
	U(r,z)=\exp \left( i f(r) \vec{n}_{R}\!\cdot\!\vec{\tau}\right) 
	\label{chiral_field} 
\end{eqnarray} 
where  
\begin{eqnarray*}
	\vec{n}_{R}=\frac{1}{1+|R(z)|^2}\left(\rm{2Re}[\it{R(z)}],
	\rm{2Im}[\it{R(z)}],\rm{1}-|\it{R(z)}|^2\right)
\end{eqnarray*}
and $R(z)$ is a rational map.

Rational maps are maps from $CP(1)$ to $CP(1)$ (equivalently, from $S^2$ 
to $S^2$) classified by winding number. It was shown in \cite{manton} 
that $B=N$ skyrmions can be well-approximated by rational maps with 
winding number $N$. The rational map with winding number $N$ possesses 
$(2N+1)$ complex parameters whose values can be determined by 
imposing the symmetry of the skyrmion. Thus, $B=3$ rational map with 
tetrahedral symmetry takes the form; 

\begin{eqnarray}
	R(z)=\frac{\sqrt{3}iz^{2}-1}{z(z^{2}-\sqrt{3}i)} 
	\label{rational_map}
\end{eqnarray}
where the complex coordinate $z$ on $CP(1)$ is identified with the 
polar coordinates on $S^2$ by $z=\tan (\theta /2){\rm e}^{i\varphi}$ 
via stereographic projection. Substituting (\ref{rational_map}) 
into (\ref{chiral_field}), one obtains the complete form of the chiral 
fields with tetrahedral symmetry and winding number 3.

Apparently, the chiral fields in (\ref{chiral_field}) takes a spherically 
symmetric form. Therefore one can apply the numerical technique developed 
for $B=1$ to find $B=3$ with tetrahedral symmetry \cite{wakamatsu}. 

Demanding that the total energy in (\ref{energy_total}) be stationary 
with respect to variation of the profile function $f(r)$,
\begin{eqnarray*}
	\frac{\delta}{\delta f(r)}E_{total}=0 \,\, ,
\end{eqnarray*}
yeilds the field equation 
\begin{eqnarray}
	S(r)\sin f(r)=P(r)\cos f(r) \label{field_equation}
\end{eqnarray}
where 
\begin{widetext}
\begin{eqnarray}
&&S(r)=N_{c}\sum_{\nu}\bigl(n_\nu\theta(\epsilon_{\nu})+{\rm sign}(\epsilon_{\nu})
{\cal N}(\epsilon_{\nu})\bigr)\langle \nu|\gamma^{0}\delta(|x|-r)|\nu\rangle \\	
&&P(r)=N_{c}\sum_{\nu}\bigl(n_\nu\theta(\epsilon_{\nu})+{\rm sign}(\epsilon_{\nu})
{\cal N}(\epsilon_{\nu})\bigr)\langle \nu|i \gamma^{0}\gamma^{5}\vec{n}_{R}
\cdot\vec{\tau}\delta(|x|-r)|\nu\rangle\,\, .
\end{eqnarray}
\end{widetext}

\begin{table}
\caption{\label{tab:helement} A schematic picture of the matrix elements  ${\bf A}(K^{\prime}
M^{\prime}, KM)$ up to $K,K^{\prime}=2$. $S$, $P_1$, $P_0$ and $P_{-1}$ refer to the elements 
coupled with $(K,M)=(0,0)$, $(1,1)$, $(1,0)$ and $(1,-1)$ respectively. Other elements are all 0.}
\begin{ruledtabular}
\begin{tabular}{cccccccccc}
      & (0 0)& (1 1)& (1 0)& (1 -1) &(2 2) & (2 1)& (2 0)& (2 -1)& (2 -2)\\
\hline
(0 0) & $S$ &       &       &       &      &      & $S$  &       & \\
(1 1) &     & $P_1$ &       &       &      &      &      & $P_1$ & \\
(1 0) &     &       & $P_0$ &       & $P_0$&      &      &       & $P_0$ \\
(1 -1)&     &       &       & $P_{-1}$ &      & $P_{-1}$&      &       & \\
(2 2) &     &       & $P_0$ &       & $P_0$&      &      &       & $P_0$ \\
(2 1) &     &       &       & $P_{-1}$ &      & $P_{-1}$&      &       & \\
(2 0) & $S$ &       &       &       &      &      & $S$  &       & \\
(2 -1)&     &  $P_1$&       &       &      &      &      &  $P_1$& \\
(2 -2)&     &       & $P_0$ &       & $P_0$&      &      &       & $P_0$ \\
\end{tabular}
\end{ruledtabular}
\end{table}
   
The procedure to obtain the self-consistent solution of Eq.(\ref{field_equation}) 
is that $1)$ solve the eigenequation in (\ref{eigen_equation}) under an assumed 
initial profile function $f_{0}(r)$, $2)$ use the resultant eigenfunctions and 
eigenvalues to calculate $S(r)$ and $P(r)$, $3)$ solve 
Eq.(\ref{field_equation}) to obtain a new profile function, $4)$ repeat $1)-3)$ 
until the self-consistency is attained.

For convenience, we shall take $f_{0}(r)=-\pi {\rm e}^{-r/X}$. 
To solve Eq.(\ref{eigen_equation}), we construct 
the trial function using the Kahana-Ripka basis \cite{kahana}; 
\begin{eqnarray}
	\Psi^{\pm}=\lim_{K_{max}\rightarrow \infty}\sum_{i=1}^{4}
	\sum_{K=0}^{K_{max}}\sum_{M=-K}^{K}\alpha_{KM}^{(i)\pm} 
	\varphi_{KM}^{(i)\pm}(r,\theta,\phi)\label{}
\end{eqnarray}
where $\Psi^{+}$ and $\Psi^{-}$ stand for parity $(-1)^{K}$ and $(-1)^{K+1}$ 
respectively, $\varphi$ is the Kahana-Ripka basis and $K$ is the grand spin
operator which is a good quantum number in the case of $B=1$ hedgehog.
The basis is discretized by imposing an appropriate boundary condition for 
the radial wavefunctions at the radius $r_{max}$ chosen to be sufficiently 
larger than the soliton size. 
And also, the basis is made finite by including only those states with the 
momentum $k$ as $k<k_{max}$. The results should be, however, independent on 
$r_{max}$ and $k_{max}$.

According to the Rayleigh-Ritz variational method \cite{bransden}, the upper 
bound of the spectrum can be obtained from the secular equation for each 
parity; 
\begin{eqnarray}
	\rm{det}\left(\bf{A}^{\pm}-\epsilon \bf{B}^{\pm}\right) = 0 
	\label{secular_equation}
\end{eqnarray}
where 
\begin{eqnarray*}
	{\bf A}^{\pm}(K^{\prime}M^{\prime}, KM)&=&\sum_{ij=1}^{4}
	\int d^{3}x \varphi_{K^{\prime}M^{\prime}}^{(i)\pm}H
	\varphi_{KM}^{(j)\pm} \\
	{\bf B}^{\pm}(K^{\prime}M^{\prime}, KM)&=&\sum_{ij=1}^{4}
	\int d^{3}x \varphi_{K^{\prime}M^{\prime}}^{(i)\pm}
	\varphi_{KM}^{(j)\pm}\,\, .
\end{eqnarray*}
For $K \rightarrow \infty$, the spectrum $\epsilon$ becomes exact. 
Eq.(\ref{secular_equation}) can be solved numerically. 

Since the chiral field in Eq.(\ref{chiral_field}) is less symmetric than 
the $B=1$ hedgehog, the hamiltonian has no grand spin symmetry. 
As a result, the states with different grand spin couple strongly and level 
splittings within the $K$ blocks occur. 
In Table~\ref{tab:helement} we present the schematic picture of the matrix elements 
${\bf A}(K^{\prime}M^{\prime}, KM)$.
Although the size of the matrix ${\bf A}(K^{\prime}M^{\prime}, KM)$ becomes quite large,  
due to the symmetry of the chiral fields, the functional space can be rearranged to reduce 
the size. Consequently, the space is divided with four blocks for each parity.  

\begin{figure}
\includegraphics{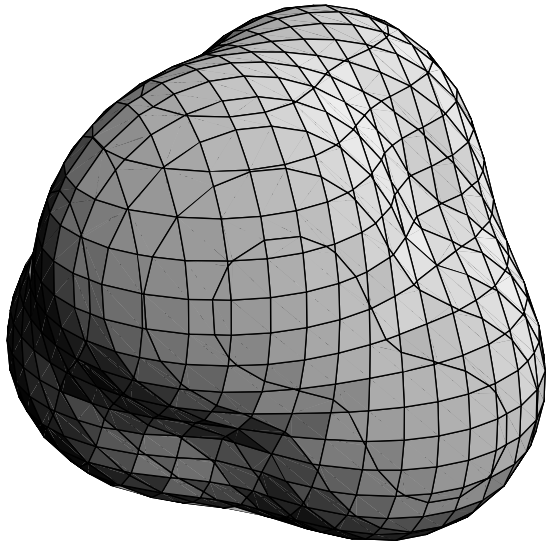}
\caption{\label{fig:density} Surface of the baryon-number density with 
$b_0=0.4\,\,{\rm fm^{-3}}$ .}
\end{figure}

\begin{figure}
\includegraphics[height=6cm,width=9cm]{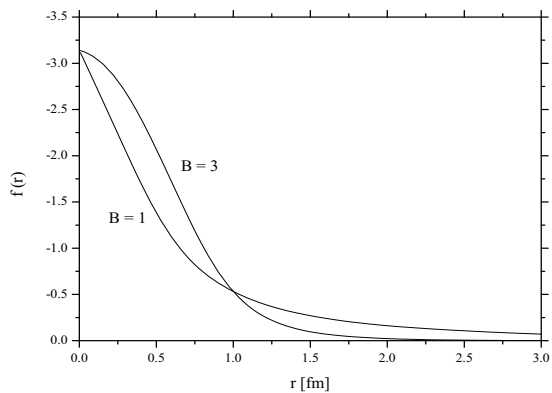}
\caption{\label{fig:profile} Profile function $f(r)$ of the rational map ansatz for $B=3$, 
and of the hedgehog ansatz for $B=1$. }
\end{figure}

Fig.~\ref{fig:spectrum} shows the spectrum of the quark orbits as a function of the 
soliton-size parameter $X$. The $P^{+}$ orbit diving into the negative 
energy region is triply degenerate. As discussed in~\cite{kahana}, baryon number of 
the soliton equals to the number of diving levels occupied by $N_{c}$ valence quarks. 
Thus putting $N_c=3$ valence quarks on each of the degenerate levels, 
one obtains the $B=3$ soliton solution. 

Fig.~\ref{fig:density} shows the corresponding baryon density. 
As can be seen, it is tetrahedral in shape. 
Therefore, we confirm that the lowest lying $B=3$ configuration is 
tetrahedrally symmetric. This result is consistent with the $B=3$ skyrmion 
obtained by Braaten {\it et.al} 
\cite{braaten}.

Fig.~\ref{fig:profile} shows the self-consistent profile function. 
For the total energy of the solution we obtain $E_{total}=3596$ MeV 
which is almost comparable to three times of the $B=1$ mass, 
and root mean square radius is $\sqrt{\langle r^2 \rangle} \sim 0.6$ fm. 
Our soliton seems to be tight object. 
This is mainly due to the missing of higher components of $K$ in our 
calculation. Their contribution becomes significant near the surface of 
the soliton and hence inclusion of the higher components will improve the 
size of the soliton. 
 
Finally, we would like to mention that our result verifies the validity 
of the rational map ansatz for the Chiral Quark Soliton Model. 
The numerical technique used here is quite general and its application 
to $B\ge 4$ will be straightforward.

\begin{center}
	{\bf Acknowledgements}
\end{center}
We are grateful to N.S.Manton for encouraging us to work on this subject 
and useful comments. We also thank V.B.Kopeliovich for suggesting
this topic.

\end{document}